\documentclass[epj]{svjour}
\usepackage{graphics}
\usepackage{bm}

\newcommand{\x}{{\bm x}}

\newcommand{\be}{\begin{equation}}
\newcommand{\ee}{\end{equation}}
\begin{document}
\title{Capillary filling for multicomponent fluid using the pseudo-potential Lattice Boltzmann method}
\author{  S. Chibbaro \inst{1}    \and L. Biferale\inst{2}  F. Diotallevi \inst{1} \and S. Succi  \inst{1}
}                     
\offprints{chibbaro@iac.rm.cnr.it}          
\institute{Istituto per le Applicazioni del
  Calcolo CNR, Viale del Policlinico 137, 00161 Roma. \and Dept. of Physics and INFN, University of Tor Vergata, Via della Ricerca Scientifica 1, 00133 Roma, Italy. }
\date{Received: date / Revised version: date}
%
\abstract{ We present a systematic study of capillary filling for a
  binary fluid by using mesoscopic a lattice Boltzmann model
  describing a diffusive
  interface moving at a given contact angle with
  respect to the walls. We compare the numerical results at changing
   the ratio the typical size of
  the capillary, $H$, and  the wettability 
  of walls. 
  Numerical results yield quantitative agreement with the
  theoretical Washburn law, provided that the channel height is sufficiently larger 
than the interface width and variations of the dynamic contact angle with the capillary number
are taken into account.
} 
\authorrunning
\titlerunning
\maketitle

\section{Introduction}
\label{intro}
The physics of capillary filling is an old problem, originating with
the pioneering works of Washburn \cite{Wash_21} and Lucas
\cite{Lucas}. It remains, however, an important subject of research
for its  relevance to  microphysics and nanophysics
\cite{degennes,washburn_rec,tas}.  Capillary filling is a
typical ``contact line'' problem, where the subtle 
non-hydrodynamic effects taking place at the contact point between liquid-gas
and solid phase allow the interface to move, pulled by capillary
forces and contrasted by viscous forces. 
As already remarked, Washburn in 1921 \cite{Wash_21} 
described theoretically the dynamics of capillary rise.
Considering also inertial effects, except the ``vena contracta'', and two fluids with the same density ($\rho_a=1, \rho_b=\rho_a$) and the same viscosity ($\mu_a=\mu_b=\mu$), the equation of motion of 
the moving front is  \cite{Ske_71}:
\be
\frac{d^2z(t)}{dt^2}+\frac{12 \mu}{H^2\rho}\frac{dz(t)}{dt}=\frac{2 \gamma cos \theta}{H\rho L}
\label{eq:wash}
\ee
where $H$ is the capillary height, $L$ its length, $\gamma$ the surface tension and $\theta$ the contact angle.
This model is obtained under the assumption that
(i) the instantaneous {\it bulk} profile is given by the Poiseuille
flow, (ii) the microscopic slip mechanism which allows  the
motion of the interface is not relevant to bulk quantities (such as the
overall position of the interface inside the channel), (iii) inlet and
outlet phenomena can be neglected (limit of infinitely long channels).
In the following, we will 
show to which extent this phenomenon can described by  a mesoscopic
Lattice-Boltzmann equation for multicomponent. The model here used is a suitable
adaptation of the Shan-Chen pseudo-potential LBE \cite{SC_93} with
hydrophobic/hydrophilic boundaries conditions, as developed in
\cite{Kan_02}. 

\section{LBE for capillary filling}
\begin{figure}
  \resizebox{0.45\textwidth}{!}{
    \includegraphics{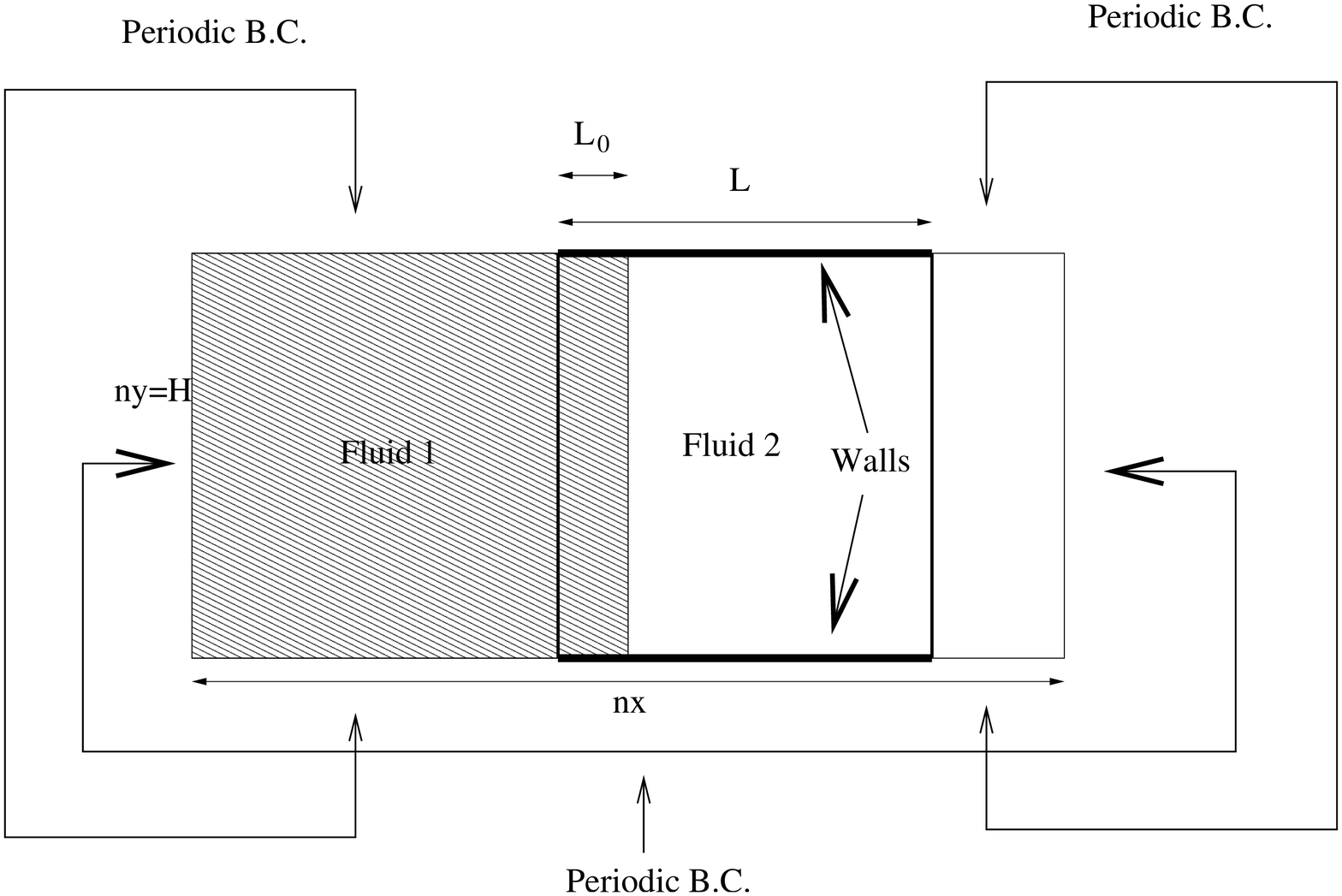}
  }
   \caption{Geometrical set-up of the numerical LBE. The two dimensional
     geometry, with length $2L$ and width $H$, is divided in two
     parts. The left part has top and bottom periodic boundary
     conditions such as to support a perfectly flat gas-liquid
     interface, mimicking a ``infinite reservoir''. In the right
     half, of length $L$, there is the true capillary:  the top and
     bottom boundary conditions are those of a solid wall, with a
     given contact angle $\theta$. Periodic boundary conditions are also imposed at the west and east sides. } 
   \label{fig:1}
\end{figure}

The relevant geometry is depicted in
fig. (\ref{fig:1}). The bottom and top surfaces are coated only in the
right half of the channel with a boundary condition imposing a given
static contact angle \cite{Kan_02}; in the left half we impose
periodic boundary conditions at top and bottom surfaces in order to
 mimic an ``infinite
reservoir''. Periodic boundary conditions are also imposed at the two
lateral sides such as to ensure total mass conservation  inside the
system. At the solid surface, bounce back boundary conditions for the particle distributions were imposed.
The conditions which allow the wetting of the surfaces will be discussed
in the following.

\subsection{LBE algorithm for multi-component flows}
Let us review the multicomponent LB model proposed by Shan and Chen \cite{SC_93}. This model allows for  distribution functions of an arbitrary number of components with different molecular mass:
\be
\label{eq:lbe}
f^k_i(\x+ {\bm c}_{i} \Delta t,t+\Delta t)-f^k_i(\x,t)=-\frac{
  \Delta t}{\tau_k} \left[f^k_i(\x,t)
  -f_i^{k(eq)}(\x,t)\right] 
\ee
where $f^k_i(\bm{x},t)$ is the kinetic probability density function
associated with a mesoscopic velocity $\bm{c}_{i}$ for the $k$th fluid,
 $\tau_k$ is a mean
collision time of the $k$th component
(with $\Delta t$ a time lapse), and $f^{k(eq)}_{i}(\x,t)$
the corresponding equilibrium function.
For a two-dimensional 9-speed LB model (D2Q9)
$f^{k(eq)}_{i}(\x,t)$ takes the following form \cite{wolf}:
\begin{eqnarray}
f^{k(eq)}_{0}&=&\alpha_kn_k-\frac{2}{3}n_k{\bf u}_k^{eq}\cdot{\bf u}_k^{eq}  \nonumber \\
f^{k(eq)}_{i}&=&\frac{(1-\alpha_k)n_k}{5}+\frac{1}{3}n_k{\bf c}_i\cdot{\bf u}_k^{eq}   \nonumber \\
&+& \frac{1}{2}n_k({\bf c}_i\cdot{\bf u}_k^{eq})^2-\frac{1}{6}n_k{\bf u}_k^{eq}\cdot{\bf u}_k^{eq} \;\;\;\textrm{for i=1$\ldots$4} \\
f^{k(eq)}_{i}&=&\frac{(1-\alpha_k)n_k}{20}+\frac{1}{12}n_k{\bf c}_i\cdot{\bf u}_k^{eq}   \nonumber \\
&+& \frac{1}{8}n_k({\bf c}_i\cdot{\bf u}_k^{eq})^2-\frac{1}{24}n_k{\bf u}_k^{eq}\cdot{\bf u}_k^{eq} \;\;\;\textrm{for i=5$\ldots$8} \nonumber \\
\end{eqnarray}
In the above equations ${\bf c}_i$'s are discrete velocities, defined as follows 
\be
{\bf c}_i= \left \{ \begin{array} {l} 0, i=0, \\
\left(cos\frac{(i-1)\pi}{2},sin\frac{(i-1)\pi}{2}\right), i=1-4 \\
\sqrt{2}\left(cos[\frac{(i-5)\pi}{2}+\frac{\pi}{4}],sin[\frac{(i-5)\pi}{2}+\frac{\pi}{4}]\right), i=5-8 
\end{array}
\right.
\ee
in the above, $\alpha_k$ is a free parameter related to the sound speed of a region of pure $k$th component according to $(c_s^k)^2=\frac{3}{5}(1-\alpha_k)$; $n_k=\sum_if^k_i$ is the number density of the $k$th component. The mass density is defined as $\rho_k=m_kn_k$, and the fluid velocity of the $k$th fluid ${\bf u}_k$ is defined through $\rho_k{\bf u}_k=m_k\sum_i{\bf c}_if_i^k$, where $m_k$ is the molecular mass
of the $k$th component.
The equilibrium velocity  ${\bf u}_k^{eq}$ is determined by the relation
\be
\rho_k{\bf u}_k^{eq}=\rho_k{\bf u}^{\prime}+\tau_k{\bf F}_k
\ee
where ${\bf u}^{\prime}$ is the common velocity of the two components.
To conserve momentum at each collision in the absence of interaction (i.e. in the case of ${\bf F}_k=0$)
${\bf u}^{\prime}$ has to satisfy the relation 
\be
{\bf u}^{\prime}=\left(\sum_i^s \frac{\rho_k{\bf u}_k}{\tau_k}\right)/
\left(\sum_i^s \frac{\rho_k}{\tau_k}\right)\;.
\ee
The interaction force between particles is the sum of
a bulk and a wall components.
The bulk force is given by
\be
\label{forcing} 
{\bf F}_{1k}({\bf x})=-\Psi_k(\x)\sum_{\x^{\prime}} \sum_{\bar{k}=1}^sG_{k \bar{k}}\Psi_{\bar{k}} (\x^{\prime})(\x^{\prime}-\x) 
\ee
where $G_{k\bar{k}}$ is symmetric and $\Psi_k$ is a function of $n_k$.
In our model, the interaction-matrix is given by
\be
G_{k \bar{k}}=\left \{ \begin{array} {l} g_{k \bar{k}}, |\x^{\prime}-\x|=1, \\
g_{k \bar{k}}/4, |\x^{\prime}-\x|=\sqrt{2}, \\
0, \textrm{otherwise}. \end{array}
\right.
\ee
where $g_{k \bar{k}}$ is the strength of the interparticle potential between 
components $k$ and $\bar{k}$. In this study, the effective number density $\Psi_k(n_k)$ is taken simply
as $\Psi_k(n_k)=n_k$. Other choices would lead to a different equation of state (see below).

At the fluid/solid interface, the wall is regarded as a phase with constant number density.
The interaction force between the fluid and wall is described as 
\be
\label{forcingw} 
{\bf F}_{2k}({\bf x})=-n_k(\x)\sum_{\x^{\prime}} g_{k w} n_w
 (\x^{\prime})(\x^{\prime}-\x)
\ee
where $n_w$ is the number density of the wall and $g_{kw}$ is the interaction strength between component $k$ and the wall. By adjusting $g_{kw}$ and  $n_w$ , different wettabilities can be obtained. 
This approach allows the definition of a static contact angle $\theta$, by introducing   a
suitable value for the wall density $n_w$ \cite{Kan_02},
which can span the range $\theta \in [0^o:180^o]$.
In particular, we have chosen $g_{1w}=0,g_{2w}=-g_{12}$ while $n_w$ is varied in order 
to adjust the wettability.
In the sequel, we choose $g_{12}=0.2$ which indicates that species $2$ is attracted  by the wall (hydrophilic), while species $1$ is neutral.
Let us note that  high values of $n_w$ are associated with hydrophilicity.

In a region of pure $k$th component, the pressure
is given by $p_k=(c_s^k)^2m_kn_k$, where $(c_s^k)^2=\frac{3}{5}(1-\alpha_k)$.
To simulate a multiple component fluid with different  densities,
we let $(c_s^k)^2m_k=c_0^2$, where $c_0^2=1/3$.
Then, the pressure of the whole fluid is given by 
$p=c_0^2\sum_kn_k+\frac{3}{2}\sum_{k,\bar{k}}g_{k,\bar{k}}\Psi_k\Psi_{\bar{k}}$, which represents a non-ideal gas law.
The viscosity is given by $\nu=\frac{1}{3}(\sum_k\beta_k\tau_k-\frac{1}{2})$,
where $\beta_k$ is the mass density concentration of the $k$th component.

The Chapman-Enskog expansion \cite{wolf} shows that the fluid mixture 
follows the Navier-Stokes equations for a single fluid:
\begin{eqnarray}
\partial_{t}\rho + \nabla \cdot (\rho {\bf u}) &=&0, \\
\rho [ \partial_{t}
    {\bm u} + ({\bm u} \cdot {\bm \nabla} ){\bm u}] &=& - {\bf \nabla}
  {P} +{\bm F} + {\bm \nabla} \cdot (\nu \rho(
                     {\bm \nabla} {\bm u}+{\bm u}{\bm \nabla}) \nonumber
\label{eq:NS}
\end{eqnarray}
where $\rho=\sum_k \rho_k$ is the total density of the fluid mixture and the whole fluid velocity ${\bf u}$ is defined by
$\rho{\bf u}= \sum_k \rho_k{\bf u}_k+\frac{1}{2}\sum_k{\bf F}_k$.

\subsection{Numerical Results}

All simulations were performed using the Shan-Chen model described above,
setting $\nu_l=\nu_g=0.167$, $\rho_l=\rho_g=1$, $g_{12}=0.2$, $\alpha =4/9$,
that is $c_s^2=\frac{1}{3}$, and the  interfacial tension is $\gamma=0.07$.
The channel length is chosen to be $L=450$.
By taking $\theta$ constant in time, a simple analytical solution of equation (\ref{eq:wash}) can be obtained:
\be
z(t)= \frac{V_{cap}H cos \theta }{6 L}t_{d}\left[ \exp(-t/t_{d}) 
+ t/t_{d} - 1 \right] +z_0,
\label{eq:analy}
\ee
where $z_0$ is the starting point of the interface at the beginning of the simulation, $t_{d}=\frac{\rho H^2}{12\mu}$ is a typical transient time and $V_{cap}=\frac{\gamma}{\mu}$ is the capillary speed.
This solution has been used to compare with simulations.
\begin{figure}
\vspace{0.25cm}
\begin{center}
  \resizebox{0.4\textwidth}{!}{
    \includegraphics{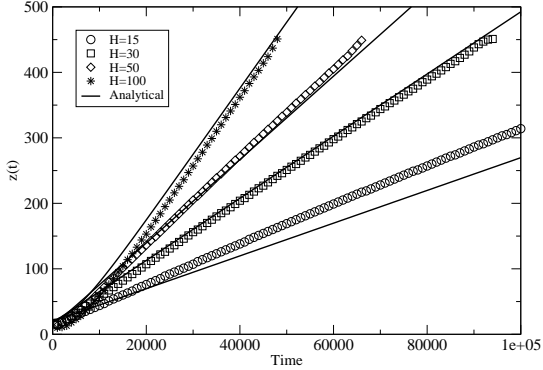}
  }
\vspace{0.4cm}
\caption{
Front displacement vs time for different channel height $H=15,30,50,100$ with their corresponding analytical solutions. The discrepancy from washburn's law is stronger for the smallest channel. The channel length is always $L=450$ except for H=100, for which $L=500$. }
\label{Fig:front_H}
\end{center}
\end{figure}
\begin{figure}
\vspace{0.25cm}
\begin{center}
  \resizebox{0.4\textwidth}{!}{
    \includegraphics{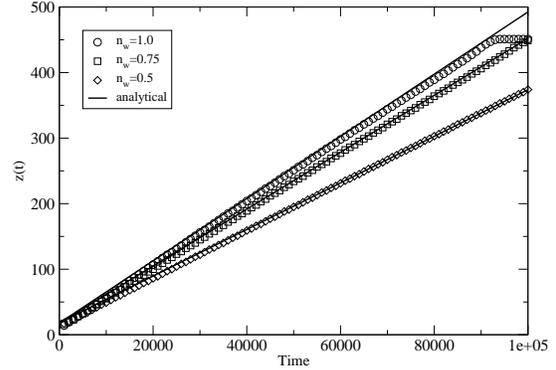}
  }
\vspace{0.4cm}
\caption{Front dynamics for different $n_w$ ($1.0,0.75,0.5$) that is for different degree of wettability. The configuration considered is with $H=30$.
The case $n_w=0.5$ is fitted by an analytical solution with $\theta=0.78$ and
the case $n_w=0.75$ with $\theta=0.52$.}
\label{Fig:frw05}
\end{center}
\end{figure}

The front displacement as a function of time is shown in Fig. \ref{Fig:front_H}
for different values of the channel height $H=15,30,70,100$,
at $n_w=1$, for which static contact angle was found to be $\theta \approx 5 $.
As expected, the velocity of the front grows with channel height.
The analytical curves are given by the solution of Eq. (\ref{eq:analy}),
where the contact angle is the dynamic one computed from numerical data.
The contact angles  computed for the four heights $15,30,50,100$ 
are respectively $0^{\circ}, 11^{\circ}, 25^{\circ}, 45^{\circ}$. 
The dynamic contact angle has been obtained directly 
as the slope of the contours of near-wall density field, and independently
through the Laplace's law,
$\Delta P= \frac{2 \gamma cos \theta}{H}$.
The latter has been chosen  for the comparison with analytical fitting curves,
because the direct computation from density contours turns out to be less precise.
Nevertheless, the values calculated in the two ways are approximately consistent.
For instance, the contact angle computed for the case $H=30$
from the direct measurements of the pressure is  $\theta \approx 12^{\circ}$ against 
the value  $\theta \approx 11^{\circ}$ computed via density contours. 
Some comments on the  front dynamics are in order..
The case of smallest channel height  does not follow the analytical solution,
showing  the finite size of the interface ($w/H~1/3$)
significantly affects the results.
On the other hand,  for a larger channel, good  agreement between
numerical and theoretical results not only holds asymptotically,
but it also extents  to the initial transient.
This is particularly true for the largest height $H=100$, where the transient time-scale 
$t_d=\frac{\rho H^2}{12\mu}$ is sufficiently long to make the exponential term in the solution (\ref{eq:analy})
important over a macroscopic time span.
The results show that the dynamic contact angles experience a strong dependence on the channel height.
In particular, in small channels, dynamic contact angles remain near their static values.
On the other hand, for large ones the discrepancy is evident.
This is due to the increasing value of the capillary number  ($Ca\sim 0.03$ for $H=100$),
since it is known that there is a correction of the dynamic contact angles 
due to finite capillary numbers.
This correction takes the form the general form $cos(\theta_d)-cos(\theta_s)=g(Ca)$.
Our results are best fitted by $g(Ca)=18~ Ca^{1.2}$,
which is in line with previous forms used in different LBE methods \cite{Roth_07,Wolf_04}

Hereafter the configuration with $H=30$ and $n_w=1.0$ will be used as a reference
for all simulations.
In figure \ref{Fig:frw05}, the front dynamics is shown for the case $n_w=1.0,~ 0.75,~ 0.5$.
As expected,  it is found that  more hydrophobic  cases correspond to smaller velocities .
The analytical solutions which fit the numerical data are obtained 
respectively with $ \theta=22^{\circ}, \theta=24^{\circ}, \theta=40^{\circ}$.
These angles are consistent  with the values
computed via Laplace's law directly from numerical data, that is $\theta=0.2,~ \theta=0.37, ~\theta=0.69$.

\begin{figure}
\vspace{0.3cm}
\begin{center}
  \resizebox{0.4\textwidth}{!}{
    \includegraphics{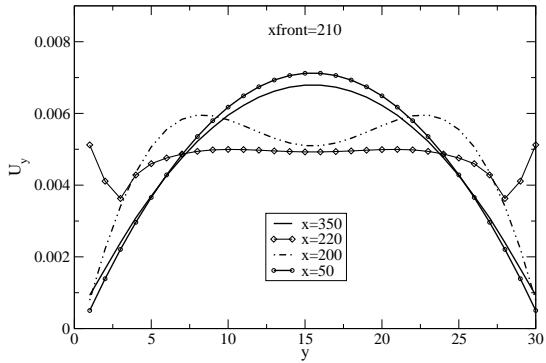}
  }
\vspace{0.4cm}
\caption{
Velocity profile $U_x(y)$  for different cuts taken at time $t=40000$ with
 the front  located at at $x\approx 210$.
One cut is taken far behind the front, $x=50$, another is far ahead  at $x=350$. 
For these cases,
 approximately the same Poiseuille parabolic flow is found.
The other two curves correspond to the velocities just ahead and behind the interface. In these case, the velocity profile is necessarily distorted in order to let the interface advance with an uniform velocity along $y$. The interface acts as an obstacle and the velocity shows a corresponding decrease (but not a recirculation) in the middle of the channel,
giving rise to a two-humped profile.} 
\label{Fig:vel_std}
\end{center}
\end{figure}

Velocity profiles taken at time $t=50000$ at different positions 
are shown in fig. \ref{Fig:vel_std}, for the standard case $H=30$, $n_w=1.0$.
Some comments are in order.
The velocity profile is parabolic everywhere except very near the interface.
This is consistent with the assumption of a parabolic (Poiseuille) velocity profile.
A small difference is present between the parabolic profile ahead and past the interface.
This is tentatively interpreted as due to the different boundary conditions applied to the fluids
($n_w=1$ for the hydrophilic invading fluid 1, and $n_w=0$ for fluid 2 ahead of the front).
This difference were found to disappear
by setting nearer values of $n_w$ for both fluids.
\begin{figure}
\begin{center}
  \resizebox{0.4\textwidth}{!}{
    \includegraphics{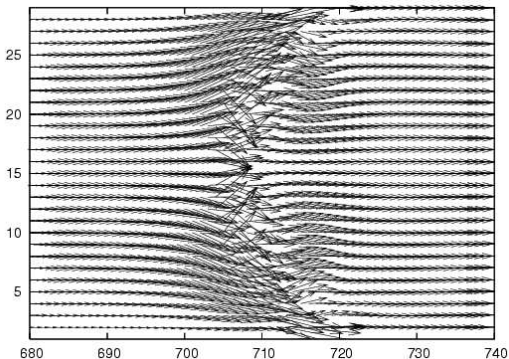}
  }
\caption{Velocity streamlines. The value of velocities are magnified by a factor $1000$. The interface is located at $x\approx 710$. Near the interface the profile is distorted and a secondary flow appears. }
\label{Fig:pres_std}
\end{center}
\end{figure}
In other terms, boundary conditions are such that the fluid after the interface is less 
slipping, with a velocity at the wall almost recovering no-slip condition.

In fig. \ref{Fig:pres_std}, velocity patterns are presented. Consistently with fig. 4, this figure shows that the flow is one-directional far from the interface, confirming the assumption of a Poiseuille flow.
Moreover, although the flow appears to be distorted near the interface to allow slippage,
no recirculation is observed at variance with other methods LBEs \cite{Fab_07,Wolf_04,Roth_07},
spurious currents are negligible.
The spikes in fig. 4 reflect the existence of a hydrodynamic singularity near the wall.
A detailed understanding of the LB dynamics in the near vicinity of this singularity remains an open issue for
future research.

\section{Conclusions}
The present study shows that Lattice Boltzmann models with
pseudo-potential energy interactions are capable of reproducing the
basic features of capillary filling for binary fluids,
as described within the Washburn
approximation.  
Moreover, it has been shown that the method is able to reproduce the 
expected front dynamics
for different degree of surface wettability,
as well as the 
correct Poiseuille velocity profile, in the whole domain, 
except for a thin region near the interface.
Quantitative agreement has been
obtained with  a sufficiently thin
interface, $w/H < 0.3$ and with two fluids at the same density.
It would  be desirable to
extend the LB scheme in such a way to achieve larger density contrasts  
and interface widths of the order of the lattice spacing
$\Delta x$ (current values are about $5 \Delta x$).
Work along these lines is  underway.

\section{Acknowledgments}
Work performed under the EC contract NMP3-CT-2006-031980 (INFLUS) and
funded  by the Consorzio COMETA within the project PI2S2 (http://www.consorzio-cometa.it).
Discussions with Dr. F. Toschi are kindly acknowledged.

\end{document}